\documentclass[epj]{svjour}
\usepackage[dvips]{graphicx}
\usepackage{amssymb,amsmath}
\usepackage{textcomp}

\begin{document}
\title{Permeability of mixed soft and hard granular material: hydrogels as drainage modifiers}
\author{E. Verneuil\inst{1}\inst{2}\thanks{\emph{Present address:} PPMD-SIMM, ESPCI and CNRS, 10 rue Vauquelin - 75005 Paris, France} \and D. J. Durian\inst{1}}                     
\institute{Department of Physics and Astronomy, University of Pennsylvania, Philadelphia, PA 19104, USA \and Complex Assemblies of Soft Matter, CNRS-Rhodia-UPenn UMI 3254, Bristol, PA 19007, USA}

\abstract{
We measure the flow of water through mixed packings of glass spheres and soft swellable hydrogel grains, at constant sample volume.  Permeability values are obtained at constant sample volume and at porosities smaller than random close packing, for different glass bead diameters $D$ and for variable gel grain diameter $d$, as controlled by the salinity of the water.  The gel content is also varied.  We find that the permeability decays exponentially in $n(D/d)^b$ where $n=N_{gel}/N_{glass}$ is the gel to glass bead number ratio and $b$ is approximately 3.  Therefore, flow properties are determined by the volume fraction of gel beads.  A simple model based on the porosity of overlapping spheres is used to account for these observations.
} %end of abstract
\maketitle
\section{Introduction}
\label{intro}

The flow of water through packings of granular materials where swellable or soft grains are added is a question relevant to a wide range of problems.
In the field, the prediction and modeling of the flows through natural soils that contain clay, which swells with water, is still subject to active research by geophysicists \cite{frenkel:1978,Levy:2005}. On the other hand, the control of both the drainage and the water retention capability of granular materials where hydrogels have been added has recently been proposed. The applications of such additives span a wide range, from baby diapers where polyacrylamide gels are added to the fiber pads \cite{Buchholz:2006}, to the sandy soils of desertified countries where hydrogels have been envisioned as a way to reduce the volume of irrigation (for a review, see \cite{kazanskii:92,Bhardwaj:2007}). In the latter case, the efficacy of superabsorbents as soil additives was shown to strongly depend from one trial to another and is still subject to controversy.\\

The properties of flows through packings of hard grains have been extensively studied, both theoretically \cite{Kozeny:1927,Carman:1937,Dullien:1979}, numerically \cite{Martys:94,Coelho:1997,Zaman:2010} and experimentally \cite{Loudon:1952,chapuis:2004}. Based on the pioneering work of Darcy on the bulk resistance of a porous medium to incompressible viscous fluid flow, the superficial fluid velocity $U=Q/S$, where $Q$ is the volume flow rate across a section $S$ of granular medium, is related to the pressure gradient $\nabla P=\Delta P/\Delta l$ in the flow direction by the following linear equation:
\begin{equation}
U=\frac{Q}{S}=\kappa\frac{\nabla P}{\eta}=KD^2\frac{\nabla P}{\eta}
\label{eq:def_perm}
\end{equation}
Here $\eta$ is the dynamic viscosity of the fluid, $\kappa$ is the permeability, with the dimension of an area, and $K$ is the intrinsic permeability, a dimensionless number characterizing the packing properties. The term $D^2$ accounts for the dependence with the inversed squared specific surface of solids in Darcy's law, and $D$ is typically a particle size. Hence, for the simplest case of spherical grains, $D$ is taken as the mean diameter \cite{Beavers:1973}.\\
Much research has been devoted to understanding the value of the permeability. For a random close packing of spheres, where the porosity is $\varepsilon\approx0.36$, the comprehensive experimental study reported in Ref.\cite{Beavers:1973} gives:
\begin{equation}
K_0=6.3 \times 10^{-4}
\label{eq:K_0}
\end{equation}
From here, the dependence of the permeability on the porosity $\varepsilon$, the average pore size, and the tortuosity was investigated for grains of various shapes: spheres \cite{weissberg:63,Beavers:1973,Martys:94,Pan:2001,Zaman:2010}, aspherical particles \cite{Coelho:1997}, sand \cite{chapuis:2004} or rocks and sintered glass beads \cite{Wong:1984}. Several quantitative relations were proposed and tested, mostly for porosities larger than the random close packing case. A review of these models can be found in \cite{PZWong:book}. Among them, the most known and used was established by Kozeny \cite{Kozeny:1927} and Carman \cite{Carman:1937} and relates the intrinsic permeability $K$ to the porosity $\varepsilon$. The Kozeny-Carman model proposes the following relationship:
\begin{equation}
K_{KC}=\frac{1}{180}\frac{\varepsilon^3}{(1-\varepsilon)^2}
\label{eq:kozeny}
\end{equation}
In this model, the pore space is seen as a bundle of paths which do not intersect each other. This is of course unrealistic since in most porous media, the pore space is highly interconnected. Still, it is in fairly good agreement with experimental studies for porosities close to random close packing, as in Refs.\cite{Loudon:1952,Beavers:1973} and, in particular, for $\varepsilon=0.36$, $K_{KC}$ in Eq.~\ref{eq:kozeny} agrees well with $K_0$ in Eq.~\ref{eq:K_0}. More recently, the percolation theory \cite{andrade:97} and self-similar modelling \cite{sen:81} were applied to model the porous media as an attempt to get more realistic descriptions.\\

When soft and/or swellable grains are added the key question of the porous volume becomes challenging. Indeed, the shape and volume of soft grains adjust to the load, making possible a porosity value smaller than the random close packing one, down to the full clogging case \cite{singh:97}. Relevant to us are the works of Weissberg \cite{weissberg:63}, and of Martys et al. \cite{Martys:94} where packings of overlapping spheres were considered, using statistical calculations in the former case, and simulation in the latter one, yielding porosities as small as 0.1. An important result of \cite{weissberg:63} is that the permeability of overlapping spheres of increasing volume is expected to decrease exponentially with the volume of the spheres. Also of interest to us is the case of sintered glass beads studied experimentally and theoretically by Wong et al. \cite{Wong:1984}. In this work, pores are seen as tubes of random size that are interconnected, and the reduction of the porous space is obtained by systematically reducing the radius of a random tube element by a fixed factor. The result is a power-law dependence of the permeability with porosity, but with an exponent that increases towards smaller porosities in an unknown way. The main result however is that the scaling behavior of the permeability of sintered glass bead samples and rocks is determined by the skewness of the pore-size distribution. The similarity between the two types of samples arises from the formation process, which reduces more the larger pores that the smaller ones.\\
It is unclear whether or not either of the above models, overlapping spheres or systematically reduced pores, is applicable to our system of mixtures of hard and soft swellable grains.  Unfortunately, very few experiments have been performed on such systems, and no systematic study can be found in the literature where every parameter was carefully controlled. Measurements of the permeability of sandy soils containing hydrogels were reported by Bhardwaj et al. \cite{Bhardwaj:2007}. The authors qualitatively show that the swelling, the size and the amount of gel added play a role on the measured permeability. However, the swelling, tuned by the ionic content of the draining solution, wasn't controlled during the course of the experiment although hydrogels release ions at the beginning of the drainage. In addition, mixtures of gels and sand tend to expand upon swelling of the gels with water. The variations in volume of the sample appear to be strongly history dependent when the mixture is in a cylindrical column, due to wall effects. This varied volume impairs the analysis because both the porous volume and the solid (gel) volume increase.\\
In this paper, to circumvent the above difficulties, we describe permeability measurements with a controlled systematic loading and swelling procedure.  In particular, novel features include: (a)
The volume of the pile is forced to remain constant, dry or wet; (b) First column filling is performed under vacuum to ensure saturation of the porous volume, and a highly salty water is first used; (c) Then, a gradual swelling of the gel is induced by flushing in less salty water; (d) Ionic contents of both the feeding solution and the collected solution after the pile are constantly monitored; (e) Rinsing with a new feeding solution is performed until the ionic content at the outlet reaches the feeding value; (f) Degased salt solutions are used, and stored in a reservoir where nitrogen is bubbled to avoid CO$_2$ bubbles nucleation. By systematically varying the amount of gel added, their swelling and the size ratio of soft and hard grains, we gradually modified the porosity of our mixtures, and were able to determine a functional form to describe the decrease in permeability upon soft (gel) particles addition from the value obtained with the initial packing of spherical hard (glass) beads.\\

\section{Experimental set-up}
\label{sec:experimental}
We use a model system of spherical glass beads and commercial hydrogels. The mixture is saturated with water of controlled salinity to set the osmotic pressure that, together with the mechanical pressure, controls the swelling of the gels \cite{rubinstein:96,vervoort:2005}. The gels are forced to swell within the porous space between the densely packed hard spheres: this way, we probe the clogging produced by the soft particles. The experiment (Fig.~\ref{fig:setup}) consists in pushing water (with salt) through a water-saturated mixture of glass beads and gels contained in a cylinder at a controlled flow rate, while measuring the pressure drop over a given height difference along the column length. The experimental details are discussed below.\\

\subsection{Materials}
\label{sec:materials}

\begin{figure}
\resizebox{0.75\columnwidth}{!}{\includegraphics{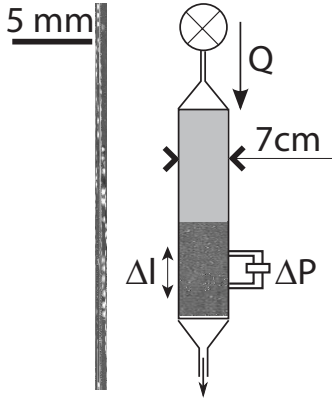}}
\caption{(Left) Photograph of the experimental system: dry gels (d$_{dry}$=0.2~mm), swollen gels (d=2~mm) and glass beads (D=0.5~mm). (Right) Schematic representation of the experimental set-up. A cylindrical column holds the mixture of gels and glass beads. Volume is kept constant by filling the extra space on top with a sponge (1-cm high) and centimer-sized balls. Water is pumped into the top of the column at a constant volume flow rate $Q$. Pressure transducer measures the pressure drop $\Delta P$ along a $\Delta l=50.8$~mm height section of the sample.}
\label{fig:setup}      
\end{figure}

\begin{figure}
\resizebox{0.75\columnwidth}{!}{\includegraphics{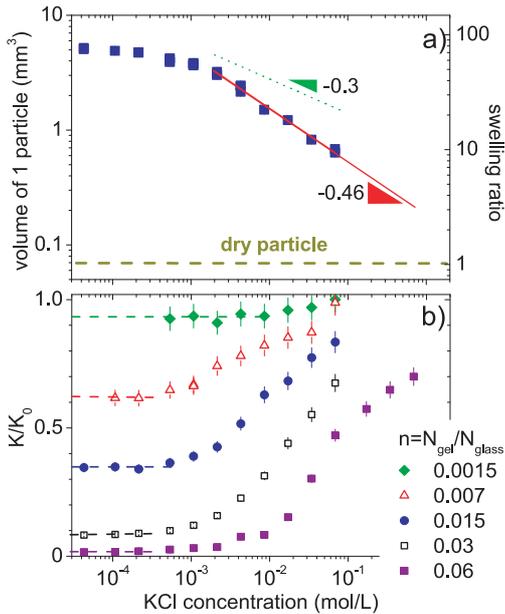}}
\caption{(a) Volume of the swollen hydrogels as a function of KCl concentration in water. Right axis: swelling ratio. At lower salt content, the swelling ratio plateaus. At higher salt content, it fits to a power law (line) with an exponent -0.46. Dotted line: theoretical prediciton from \cite{rubinstein:96}. (b) Mixtures of 0.5 mm glass beads and hydrogels at various gel to bead number ratios: Normalized permeability $K/K_0$ as a function of KCl concentration in the draining water. The dashed lines indicate the plateau value at low salt content.}
\label{fig:salt}      
\end{figure}

The glass beads (Potters Industries, PA) are spherical. Three bead diameters $D$ were used: 0.5, 1, and 2 mm. The size distributions are measured using an automated sizer (CamSizer) and fit to a Gaussian curve with a width of order 100~$\mu$m and a mean diameter $D$. We also measured the average mass of one glass bead for each size by weighing several hundreds of carefully counted beads. Prior to being used, the glass beads are treated so as to fully hydrate their surface. They are first burnt in a furnace at 500C for 72 hours and then acid-washed in a 1M HCl bath for 1 hour. The beads are then thoroughly rinsed with pure deionized water and dried in a vacuum oven at 110C.\\
The gel particles were kindly provided by Degussa Inc. (Stockosorb SW) and are shown in Figure~\ref{fig:setup}. They consist of a cross-linked copolymer of potassium acrylate and acrylamide (50/50). As a result of the manufacturing process that produces the particles by grinding a bulk gel and then sieving, their shape is faceted and random. The dry particles are further sieved between 212 and 300-micrometer sized meshes. The swelling ratio of these particles is measured for KCl solutions of increasing salt concentrations as follows. Batches of 100 particles were carefully counted, then soaked into a solution of given salinity in a tared weighing dish. After an hour, the excess of liquid was sucked up with a pipette and a paper wipe. The mass of the swollen particles is then measured with a precision scale. The process is repeated for every (higher) salt concentrations. Three different batches of 100 particles were used and we eventually obtain the average mass of one gel particle at increasing salt contents. We also measured the average mass of one dry gel particle for over 300 specimens: $m=14.8 ~\pm 0.5 ~\mu g$. The swelling ratio of the gel is defined as the ratio of the swollen to the dry mass. The average volume $v_{gel}$ is calculated from the mass using the density of the salt solution as a rough approximate for the particle density. The results are shown in Figure~\ref{fig:salt}-a). At low salt concentration, the swelling ratio is insensitive to the salt concentration and plateaus: the cross-linked polymer network is swollen to its maximum extent. For [KCL] values larger than $2\times 10^{-4}$~mol.L$^{-1}$, it decreases and the data can be fit to a power law with an exponent -0.46. This result is in good agreement with other experimental data reported in the literature \cite{rubinstein:96} for high salt concentrations that yield an exponent ranging between -0.3 (dotted line, theoretical value) and -0.6. It can be noted that in the range of salt content used here the volume of the gel particles changes by one order of magnitude.\\

\subsection{Procedures}
\label{sec:procedures}

The permeability measurement apparatus, shown in Figure \ref{fig:setup}, consists in a custom-made cylindrical column in polycarbonate (height 30.5 cm, inner diameter 7 cm, McMaster-Carr). It is closed at the bottom end by a brass sieve (mesh size 90 micrometers) that is attached to a funnel. The samples poured in the experimental column are prepared by mixing small amounts of glass beads (typically 25 to 100 g) and dry gel particles to ensure a good homogeneity of the final mixture, to a total bead mass of 700 g. Using this bead mass and measuring the volume occupied by the dry glass beads in the column, we checked that the beads arrange themselves in a statistical packing of porosity $\varepsilon$=0.38$\pm 0.04$, slightly higher than the usual figure for the random close pack porosity of 0.36. Note that this dry porosity based on the glass beads, only, is unchanged by the addition of even the larger amounts of dry gel. Hence, the initial packing of the spheres is not affected by the presence of dry gels. For data analysis, we use the average masses of individual gels and of individual glass beads to convert the gel to glass bead mass fraction that we use in practice into a gel to glass bead number ratio $n$ defined as:
\begin{equation}
n=N_{gel}/N_{glass}
\label{eq:def_n}
\end{equation}
The upper end of the column is sealed by a plug connected either to a vacuum pump (filling step) or to a gear pump (measurement step). First, the column is filled with a KCl solution at high concentration (0.07 mol.L$^{-1}$) under vacuum and from the bottom. This procedure allows for the saturation of the pores with the solution. The solutions used are prepared with deionized water, and then degased under vaccum to minimize the nucleation of CO$_2$ pockets within the sample. Nitrogen is bubbled in the water reservoir during the entire experiment (filling and measurement) to prevent the redissolution of CO$_2$ in water. During the filling, the volume of the glass bead and gel mixture is kept constant by adding a sponge and marbles on the top of the sample, and by using a salty solution to minimize the gel volume expansion. Consequently, the gels swell only within the porous space available between the glass beads.\\
The upper plug is then connected to a gear pump (Micropump Inc.). We use two different pump heads depending on the range of flow rates desired: 0.092mL/revolution for flow rates $Q$ ranging from 15 to 330 mL/min and 0.017 mL/min for $Q$ between 3 and 60 mL/min. On the side of the column, along its height, two ports are drilled through the column wall at a distance $\Delta l=50.8$~mm, to allow the connection of a wet-wet differential pressure transducer (26PCA, Omega). The pressure sensor is calibrated by connecting a water column of controlled height. It has a typical sensivity of 3.5~mV/kPa. The range of $Q$ is chosen so as to probe a pressure drop variation of one order of magnitude. We carefully checked that the pressure drop $\Delta P$ linearly increases with the flow rate $Q$ for every measurement performed: this shows that all flows are in the viscous regime. The pressure drop $\Delta P$ over $\Delta l$ is used to compute the permeability $\kappa$ and the intrinsic permeability $K$ defined by equation~\ref{eq:def_perm}. Here, the intrinsic permeability $K$ is computed using the glass bead diameter $D$ as the typical size that normalizes $\kappa$.
For every change in salinity of the draining solution, the mixture is rinsed until the ionic content at the lower end of the column, measured with an electric conductivity-meter, reaches the reservoir value.\\

\begin{figure*}
\resizebox{2\columnwidth}{!}
{\includegraphics{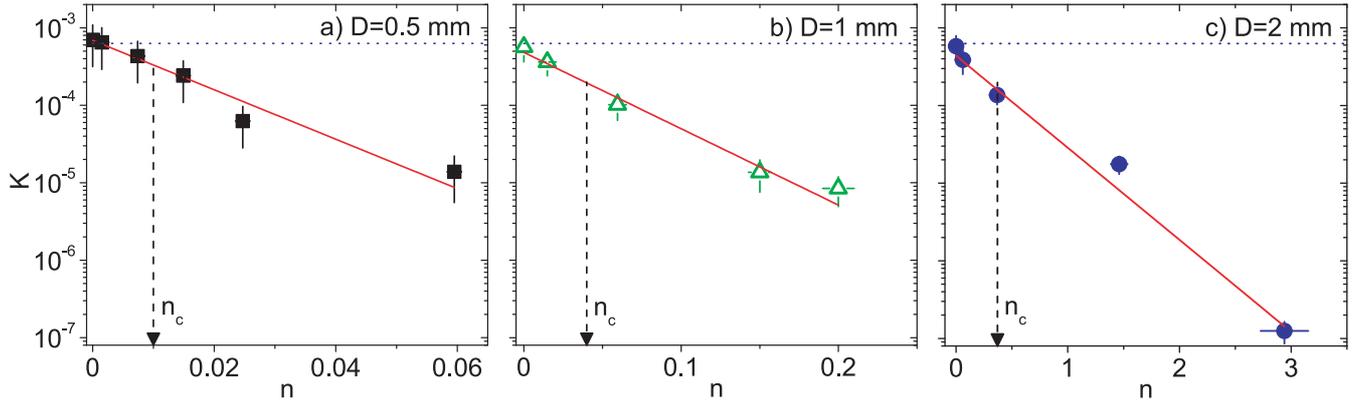}}
%\vspace*{4cm}
\caption{Low salt plateau value of the intrinsic permeability $K$ as a function of the gel to glass beads number ratio $n$ for three different glass bead diameters: a) 0.5 mm b) 1mm c) 2 mm. All the data can be fit to an exponential decay function (full lines) with a characteristic number ratio $n_c$ shown by dashed lines. The dotted line indicates the value $K_0$ in Eq.~\ref{eq:K_0}.}
\label{fig:3}      
\end{figure*}

\section{Results}
\label{sec:results}

First, we focus on the pure glass bead systems as a way to test our set-up. The permeability of the glass beads without gel, $K_0$, is plotted in Figure~\ref{fig:3} for the three $D$ values at the abscissa, gel to bead number ratio $n$=0. The horizontal dotted lines indicate the value $K_0=6.3 \times 10^{-4}$ given by Eq.\ref{eq:K_0}. Averaged together, our three measurements at $n=0$ give $K_0=(6.1\pm 0.7) \times 10^{-4}$, in good agreement with Eq.~\ref{eq:K_0}. This confirms the integrity of our procedures and indicates that the size of the sample is large enough to neglect wall and edge effects.\\

We now turn to the gel and glass bead mixed systems.\\
In a first series of experiments, we used the salinity of the draining solution to control the flow properties of a packing of 0.5-millimeter glass beads with gels. The series is performed with solutions of decreasing salinity. The permeabilities obtained are shown in Figure~\ref{fig:salt}-b as a function of the KCl concentration in the solution. For every gel content tested, we observe that the permeability plateaus at small salt concentration and then increases. The cross-over between these two regimes occurs at a salinity comparable to the salinity where the size of a single gel particle starts decreasing (Fig.~\ref{fig:salt}-a). For [KCl] concentrations larger than $5\times 10^{-4}$~mol.L$^{-1}$, the permeability follows the same trend but inverted as the swelling ratio of one gel particle. As an example, a sample with a mass fraction of 0.4\% (or number ratio $n=0.06$) has a permeability that spans the range 2\% to 70\% when increasing the KCl concentration from $10^{-3}$ to 1 mol.L$^{-1}$. This shows that salinity stands as an efficient external trigger for the permeability: the typical size of the gel particles appears to be a characteristic size of the problem. On the other hand, it is important to rinse the granular pile before acquiring any data because of excess ions in the dry gels. Omitting to take into account the excess ions in the dry particles leads to misinterpretations of measurements of the permeability with these gels, and to time-dependent measurements as in \cite{Bhardwaj:2007}.\\

\begin{figure}
\resizebox{0.75\columnwidth}{!}{\includegraphics{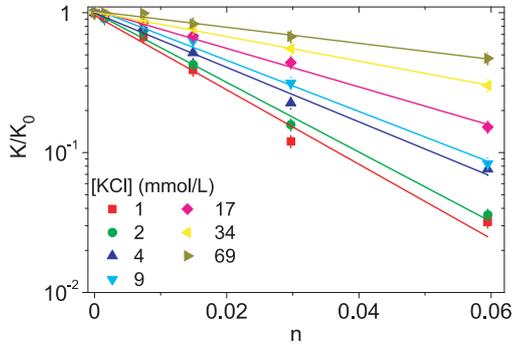}}
\caption{Mixtures of 0.5~mm glass beads and gels. Permeability data from Fig.~\ref{fig:salt} versus gel to glass number ratio. Each curve corresponds to a given salt concentration. Data fit to an exponential decay function (lines) with a characteristic number ratio $n_c$ reported in Fig.\ref{fig:nc_Dd} as diamonds for every KCl concentration.}
\label{fig:4}      
\end{figure}

In addition, we verify the obvious fact that larger gel contents yield smaller conductivities. The five curves presented in Figure~\ref{fig:salt}-b correspond to mass fractions of 0.4\%, 0.2\%, 0.1\%, 0.05\% and 0.01\% from bottom to top. Let us now focus on the low salinity case. This series at constant salt is meant to isolate the effect of the size ratio from other effects such as the elastic properties of the gels that are modified together with the gel size when salt is added \cite{rubinstein:96}. The plateau value at small salinity is measured for every gel content used and is plotted as a function of the gel to glass bead number ratio $n$ in Figure~\ref{fig:3}-a. The same data sets are collected for two more glass bead sizes (1 and 2-mm) and are shown in Figs.~\ref{fig:3}-b,c. Interestingly, for all bead sizes, the permeability decreases exponentially with the gel to glass bead number ratio. The data sets fit to an exponential decay function with two fitting parameters: a characteristic number ratio $n_c$ and the permeability at zero gel content $K_0$:
\begin{equation}
K=K_0e^{-\frac{n}{n_c}}
\label{eq:K_n_nc}
\end{equation}
with $K_0=(6.1\pm 0.7) \times 10^{-4}$.\\
We use equation \ref{eq:K_n_nc} to analyze our measurements of permeability as a full function of salinity, for all gel mixtures with $D=0.5$~mm glass beads. For every salt content, the conductivity, normalized by the reference without gel, $K_0$, is plotted as function of $n$ in Figure~\ref{fig:4}. All the data can be fit to the exponential decay function given by Eq.\ref{eq:K_n_nc} with a characteristic number ratio $n_c$.\\

\begin{figure}
\resizebox{0.75\columnwidth}{!}{\includegraphics{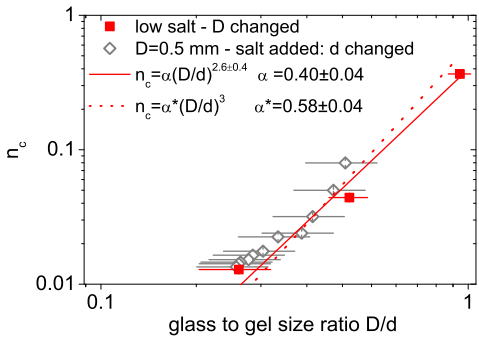}}
\caption{Characteristic number ratio $n_c$ as a function of the glass bead to gel size ratio $D/d$. Squares: $D$ is changed, $d$ fixed. $n_c$ fits to $\alpha(D/d)^b$ (full line) with an exponent $b$ close to 3 and $\alpha$ of order 1. Diamonds: $D$=0.5 mm, $d$ is decreased by KCl addition. All the data fit to $\alpha^*(D/d)^3$ (dotted line).}
\label{fig:nc_Dd}      
\end{figure}

We now turn to the analysis of the characteristic number ratio $n_c$ measured in Figures \ref{fig:3} and \ref{fig:4}. It was earlier found that the gel volume is a characteristic parameter of the problem (see Figure~\ref{fig:salt}). The two series of experiments, with different glass bead sizes and with gel sizes decreased by salt addition, can be seen in terms of a variation of the size ratio of the two types of particles. For the glass bead size, we take their diameter $D$ while we define a typical gel size $d$ at a given salt concentration as $v_{gel}=\frac{\pi}{6} d^3$, with $v_{gel}$, the swollen volume of gel particles taken from Fig.~\ref{fig:salt}-a. Hence, we plot $n_c$ as a function of the glass to gel size ratio $D/d$. We find that the dependency of $n_c$ with $D/d$ is consistent in both situations, namely when this ratio is varied either by decreasing the gel size (salt added, diamonds) or by increasing the glass bead size (square dots). Moreover, the data can be fit to a power law: $n_c=\alpha(D/d)^b$ with $\alpha \sim 0.40 \pm 0.04$ and $b\sim 2.6 \pm 0.4$ (full line in Fig.~\ref{fig:nc_Dd}).
Interestingly, the exponent $b$ is close to $3$. In Figure~\ref{fig:nc_Dd}, the dotted line corresponds to a power law with an exponent 3 and an adjustable prefactor. It nicely fits the data within our experimental incertainty with a prefactor $\alpha^*=0.58\pm 0.04$. Hence the $(D/d)^{b}$ dependence can be seen as the ratio between the volumes of both types of particles, leading to:
\begin{equation}
n_c=\alpha^* \left(\frac{D}{d}\right)^3
\end{equation}
Therefore, the argument in the exponential function can be recast as a function of the gel to total solid volume fraction $\phi$:
\begin{equation}
\phi=\frac{nd^3}{nd^3+D^3}
\label{eq:def_phi}
\end{equation}

\begin{figure*}
\resizebox{1.75\columnwidth}{!}
{\includegraphics{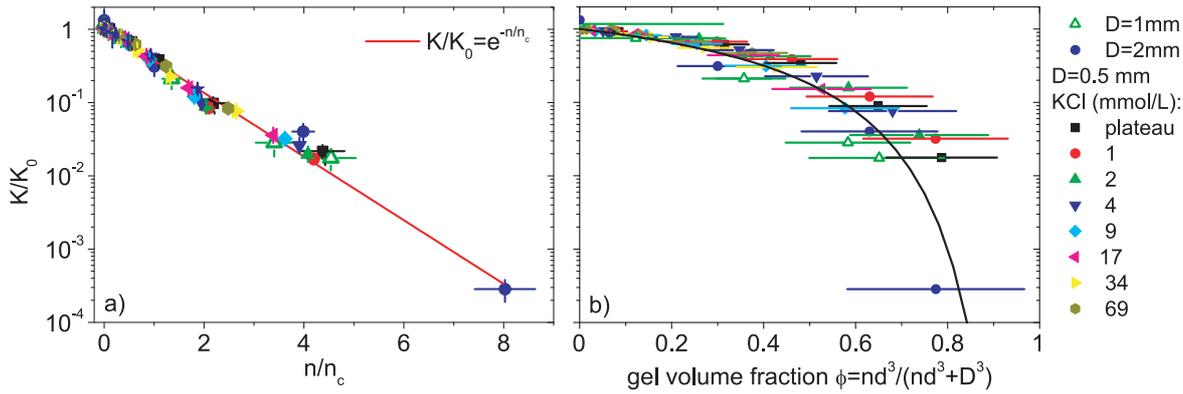}}
%\vspace*{4cm}
\caption{a) Permeability data for different glass bead sizes and different gel particle sizes collapse on a master curve providing that $K$ is normalized by its value without gel $K_0$, and the number ratio $n$ by its characteristic value $n_c$ obtained from the exponential fit. Full line: exponential decay function (no adjustable parameters).(b) Normalized permeability as a function of gel to total solid volume fraction $\phi=n(\frac{d^3}{nd^3+D^3})$ for the three glass bead diameters $D$ studied and upon gel shrinking by KCl addition into 0.5-mm sized glass beads systems. Line: exponential decay function $K/K_0=e^{-\frac{\phi}{\alpha^*(1-\phi)}}$ with $\alpha^*=0.58$.}
\label{fig:6}      
\end{figure*}

Altogether, we find that the normalized permeability $K/K_0$ plotted as a function of the normalized number ratio $n/n_c$ collapses on a single master curve shown in Figure~\ref{fig:6}-a. This master curve can be used to predict the drainage reduction provided that the characteristic number ratio $n_c$ is calibrated as a function of the glass to gel size ratio $D/d$. The calibration shown on Fig.~\ref{fig:nc_Dd} allows for the creation of another master curve where $\phi$ is the unique parameter, plotted in Figure~\ref{fig:6}-b.

\section{Discussion}
\label{sec:discuss}

%\ A closer look at Figure~\ref{fig:nc_Dd} shows that the number of gel particles required to block the flow by a given amount becomes slightly higher towards the larger $D/d$ ratios when salt is used as an external trigger. This is consistent with the fact that salt not only induces a size decrease, but also an increase in the elastic modulus. Contacts made on harder elastic material yields smaller contacts areas and larger pores left open. More gels are thus needed when the consitutive material of the soft particles is stiffer.\\
Our results show that the permeability of mixtures of soft and hard grains can be described by the general law
\begin{equation}
K=K_0 e^{-\frac{N_{gel}/N_{glass}}{\alpha^*(D/d)^{3}}}=K_0 e^{-\frac{\phi}{\alpha^*(1-\phi)}}
\label{eq:HC_Dd}
\end{equation}
Here $\alpha^*=0.58\pm 0.04$ is a numerical constant acounting for the gel softness (or equivalently, the salt content), $K_0$ is the intrinsic permeability of the glass beads pile given by Eq.~\ref{eq:K_0}, and $\phi$ is the gel to total solid volume fraction.\\

In the following, we derive simple expressions to account for our results, as summarized by Eq.\ref{eq:HC_Dd}. Because our gels are so soft, we first assume that their effect on the packing is to close some pores off while pores that aren't affected by a gel particle remain with an unchanged conducting capability. Hence, the number of open pores is decreased but the size and shape of these open pores remain unchanged.
To estimate the new porosity, we use the derivation by Weissberg \cite{weissberg:63} for overlapping spheres. Here we reproduce the main ingredients. The porous volume is made of the points in the space that are not contained by any sphere. The porosity is thus regarded as the probability that a random point is not contained by any sphere, or, equivalently, as the probability that the portion of the space which lies within a distance equal to the particle radius from the random point contains no sphere centers. First, for overlapping, monodispersed spheres of volume $v$, we derive the probability of placing $N$ centers at random in a large finite volume $V$ in such a way that a smaller volume equal to $v$ contains no centers. In this case, each random placement of a center is independent of the positions of the other centers, this probability is
$[(V-v)/V]^N = [1-v/V]^N$. For a larger and larger system where $N$ tends to infinity, holding the sphere volume $v$ and the number density $N/V$ fixed, the probability that the volume $v$ contains no centers becomes: $\exp[-N \frac{v}{V}]$.\\
Now, we turn to the two-size problem. We have glass beads of size $D$ that are non-overlapping, and gel overlapping particles of volume $v\equiv_{gel}=\frac{\pi}{6}d^3$. The probability for a point to be in the porous space is the product of the two probabilities given by each type of particles. Because the glass beads don't overlap, the corresponding term is simply the initial porosity $\varepsilon$. The gel beads are assumed to be spherical. The corresponding term is then: $\exp[-\frac{\pi}{6} d^3 N_{gel}/V]$.\\
Altogether, we find that the porosity of a composite system with hard glass spheres, and soft, overlapping gel particles is given by: $\varepsilon~\exp[-\frac{\pi}{6} d^3 N_{gel}/V]$.\\
From this, the derivation of the conductivity is made following the method developped by Carman \cite{Carman:1937}. Here, we assume that the pores already present in the glass sphere pile are either closed by the gel or left open and unchanged. Thus, the conductivity is simply the product of the initial conductivity $K_0$ scaled down by the porosity factor, yielding:
\begin{equation}
K=K_0 e^{-v_{gel} N_{gel}/V}
\end{equation}
This expression can be recast using the experimental variables used to fit our data in Fig.\ref{fig:6}. The total sample volume $V$ can be calculated using the glass beads volume since dry gels occupy a small enough volume for it to be neglected and our experimental procedure prevents the sample from swelling. Then, $V=N_{glass}v_{glass}/(1-\varepsilon)$, where $v_{glass}=\frac{\pi}{6} D^3$ is the volume of one glass sphere. Hence:
\begin{equation}
K=K_0 e^{-(1-\varepsilon)\frac{v_{gel} N_{gel}}{v_{glass}N_{glass}}}
\end{equation}

Our calculation, however rough it is, yields an exponential decaying function of the total gel volume to the total glass volume, in remarkably good agreement with our experimental results. The derivation first assumes a spherical and uniform shape for the gel beads which is quite a strong assumption. Second, modelling the gels as overlapping spheres also assumes a loss of volume when gel is squeezed between the glass spheres. Refering to studies of highly swollen gels under compression \cite{vervoort:2005}, we know that under large strains, gels release solvent and deswell. Here, the size of the pores left between glass spheres are about ten times smaller than the gel size when unstrained, so the strain applied to the gels is high and deswelling probably occurs. Also, with our procedure that prevents the pile from swelling, the pile is submitted to a load that can be as high as needed for the distance between glass beads to remain constant. Hence, the effect of an increased osmotic pressure upon salt addition is equivalent to an increased mechanical pressure yielding deswelling, although salt addition leads to both a decrease in size and a increase in elastic modulus \cite{vervoort:2005}. This explains why experiments made with salty solutions could be analyzed in terms of size (Fig.\ref{fig:nc_Dd}) without taking into acount the increased stiffness of the gel. Moreover, our model assumes that gel particles totally close some pores while others remain unchanged. In the framework developped by Carman, pores are seen as a bundle of parallel tubes of uniform diameter. The above assumption means that gels do not partially invade a tube. This hypothesis is supported by the large size ratio between gels and pores and the high softness of the gel material (Young's modulus of order 1 to 10~kPa). Hence, a gel will swell into any neighboring pore "tube" so as to occlude it and to get in contact with the next grain. 
Finally, the model neglects excluded volume in assuming that the addition of a new gel does not depend on the gel beads that are already present. This would increase the slope of the decaying exponential behavior. Looking at the argument of the exponential function, our experimental results yield a prefactor $1/\alpha*=1.72$ while our model predicts a prefactor $(1-\varepsilon)=0.64$. Among other effects, the discrepancy could in part come from the excluded volume effect mentionned above.

%\$\rho=0.998$~kg.m$^{-3}$, $\eta=0.8907$~mPa.s,

\section{Conclusion}
\label{sec:ccl}

To determine the properties of flows through a mixed packing of soft and hard particles, we presented a series of measurements of the hydraulic conductivity of a pile of glass spheres in which a controlled amount of hydrogel particles was added. The volume of the pile was forced to remain at the size of the glass bead pile without gels. Care was taken to fully saturate the porous volume with water at all time, and to precisely control the ionic content of the water flowing in the pile. Hence, the efficient volume available for conduction and the solid (gel) volume were known at all time, thus improving upon prior work. We find that the conductivity of the pile is equal to its value without gels scaled down by an exponentially decaying function of the volume fraction of gel presented in Eq.~\ref{eq:HC_Dd}. This result can be explained by modelling the gel particles as overlapping spheres. The porosity of the assembly of hard non-overlapping glass spheres and soft overlapping gel particles was derived, and from there, the conductivity of the pile was deduced assuming that pores are either open or fully occluded by gel. This simple derivation holds because our gels are soft and large enough compare to the glass beads.\\
Turning to the efficacy of gels as drainage modifiers, our results allows for the determination of the right amount of gel to add to get a given reduction in conductivity. This work could be continued by taking into acount the swelling of the pile. The experimental device should be carefully designed so as to carefully control the volume expansion of the pile upon gel swelling, and to make it independent of the history of the sample and of wall effects.\\

\section{Ackowledgements}
We thank Douglas J. Jerolmack for fruitful discussions and for use of his Camsizer. We than Jean-Christophe Castaing for drawing our attention to this problem, and for his insights on the results. We are also grateful to Degussa Inc. for providing us with hydrogel samples. This work was partially supported by the National Science Foundation through grants MRSEC/DMR05-20020 and DMR-0704147.

%\section{Optional figure}
%Could be between figures 1 and 2.
%\begin{figure}
%\resizebox{0.75\columnwidth}{!}{\includegraphics{figure_8.eps}}
%\caption{Permeability $\kappa_0$ as a function of the glass bead diameter $D$. As expected, $\kappa_0$ scales as $D^2$ (line) and the prefactor gives an intrinsic permeability $K_0$ close to the value found in the literature (Eq.~\ref{eq:K_0}).}
%\label{fig:K0}      
%\end{figure}

% BibTeX users please use
 \bibliographystyle{unsrt}
 \bibliography{biblio}
%\begin{thebibliography}{}
%Author, \textit{Book title} (Publisher, place year) page numbers
% etc
%\end{thebibliography}

\end{document}